\documentclass{article}

\usepackage{PRIMEarxiv}

\usepackage[utf8]{inputenc} 
\usepackage[T1]{fontenc}    
\usepackage{hyperref}       
\usepackage{url}            
\usepackage{booktabs}       
\usepackage{amsfonts}       
\usepackage{nicefrac}       
\usepackage{microtype}      
\usepackage{graphicx}       
\usepackage{svg}
\usepackage{float}
\usepackage{subcaption}
\usepackage{textgreek}
\usepackage{amsmath} 

\graphicspath{{media/}}

\title{Optical vortex trajectories as probes for wavefront aberrations}

\author{
  Aleksandra K. Korzeniewska, Magdalena \L{}ukowicz, Kamil Kalinowski, Karolina Gemza, Mateusz Szatkowski\\
  Wroc\l{}aw University of Science and Technology\\
  Department of Optics and Photonics\\
Wybrzeże Wyspiańskiego 27, 50-370 Wrocław, Poland\\
  \texttt{mateusz.szatkowski@pwr.edu.pl} 
}

\begin{document}
\maketitle

\begin{abstract}
Phase singularities, due to their high sensitivity to phase disturbances, are a promising tool for wavefront retrieval. Several methods have been proposed to exploit this property, one of which analyzes their trajectories—the paths that singular points follow when shifted off-axis. Nevertheless, the relations between primary aberrations and trajectories remain unexploited. In this work, we aimed to address this gap by investigating how distinct aberrations influence vortex trajectory behavior. We performed numerical simulations of vortex trajectories under manually introduced aberrations and proposed metrics to describe their relationship. Our results show that defocus, coma, and astigmatism each produce a unique trajectory response, allowing differentiation between aberration types present in the beam. We proposed and experimentally validated an autofocusing algorithm that leverages the trajectory shape to identify the back focal plane of the optical system. This work presents a comprehensive study of optical vortex trajectories to further advance wavefront sensing based on phase singularities.
\end{abstract}


\section{Introduction}

The term phase singularity, also known as an optical vortex, refers to a point in an optical field where the phase is undetermined, corresponding to a zero intensity \cite{Nye1974}. In the close vicinity of such a point, the phase changes $m$ times by $2\pi$, where $m$ denotes the topological charge \cite{Basistiy1995, Bazhenov1992}. Its fundamental example is a beam carrying a single vortex at its center, carrying the orbital angular momentum (OAM) $L$ proportional to its topological charge, $L=m \hbar$ , where $\hbar$ is the reduced Planck's constant. While OAM is conserved during propagation, the position of the singularity is sensitive to external perturbations. The dynamics of these topological features can reveal subtle changes in the optical field. 

Early studies showed that increasing the topological charge leads to vortex splitting under aberrations, with high-order vortices ($|m|>1$) breaking into single charge ones \cite{Freund1999, Dennis2012}. This became the foundation for linking singular point behavior to optical aberrations.

In \cite{Roux2003}, the analytical model for computing the evolution of a decentered vortex has been introduced, followed by the work analyzing the evolution of a vortex dipole - pair of optical vortices with topological charges of $m=1$ and $m=-1$ \cite{Roux2004}. Further, the evolution of singular points has been studied both analytically and experimentally, leading to the concept of singular skeleton - a continuous trace of singularities in space. 

Various research addressed this topic and reported on its dynamics under edge-diffraction \cite{Bekshaev2017} or by the double-phase-ramp converter consisting of two alternated optical wedges \cite{Khoroshun2017, Khoroshun2020}. Similarly, the behavior of singular point traces has been studied when diffracted by the element possessing rotational symmetry of finite order \cite{Ferrando2013}, linking the shape of the trace with the diffractive object, aiming to establish a controllable way to structure the evolution of singular points.

Researchers studied more complex singular structures such as vortex knots \cite{Leach2005, Dennis2010}, which exhibit topological robustness \cite{Dehghan2023} and potential for high-capacity information encoding \cite{Larocque2020, Kong2022}. Other works examined vortex lattices \cite{Ortega2019, Karen_Volke_2022} and the dynamics of multiple vortex pairs, revealing mechanisms like pair annihilation in structured fields \cite{Ferrando2023}. It has also been shown that the position of a singular point could be used for object scanning in microscopy setups \cite{Augustyniak_2012, P-Masajada_2018}. Intentional shift of the vortex-generating element off-axis produces a singular point trajectory in the observation plane \cite{Plocinniczak_2016}, that is sensitive to wavefront distortions.

The light's spatial and phase properties, including singular point position, can be treated as independent degrees of freedom \cite{He2022} while simultaneously promising high metrological sensitivity due to their super-oscillatory nature \cite{Berry_2019, Gbur2018, Chen2019}.

Building on these concepts, vortex trajectory behavior could lead to the development of a simple aberrometer. Results in \cite{Szatkowski_2019} demonstrated that vortex trajectories are an objective method to assess beam quality. Following that approach, any optical system equipped with a spatial light modulator or digital micromirror device can introduce a phase singularity, shift it off-axis, and evaluate the beam. 

However, no detailed numerical study has been proposed to examine how particular aberrations affect vortex trajectories, and neither model has linked trajectory features to specific aberrations quantitatively. In this work, we address this gap by performing numerical simulations to analyze the effects of defocus, astigmatism, and coma on optical vortex trajectories. We introduce new metrics to characterize these aberrations and explore their potential applications. While our primary objective is to advance optical vortex aberrometry, we also show how vortex trajectories can be leveraged for autofocusing, offering an easily implementable solution, which we discuss in detail.

\section{Numerical simulations}
\subsection{Simulation Methodology}
To calculate the behavior of the optical vortex trajectory under controlled aberration, the proposed numerical model (Figure \ref{fig:setup_schematic}a) is implemented using VirtualLab Fusion. The model consists of the following optical components:

\begin{figure}[htbp]
\centering
\includegraphics[width=280 pt]{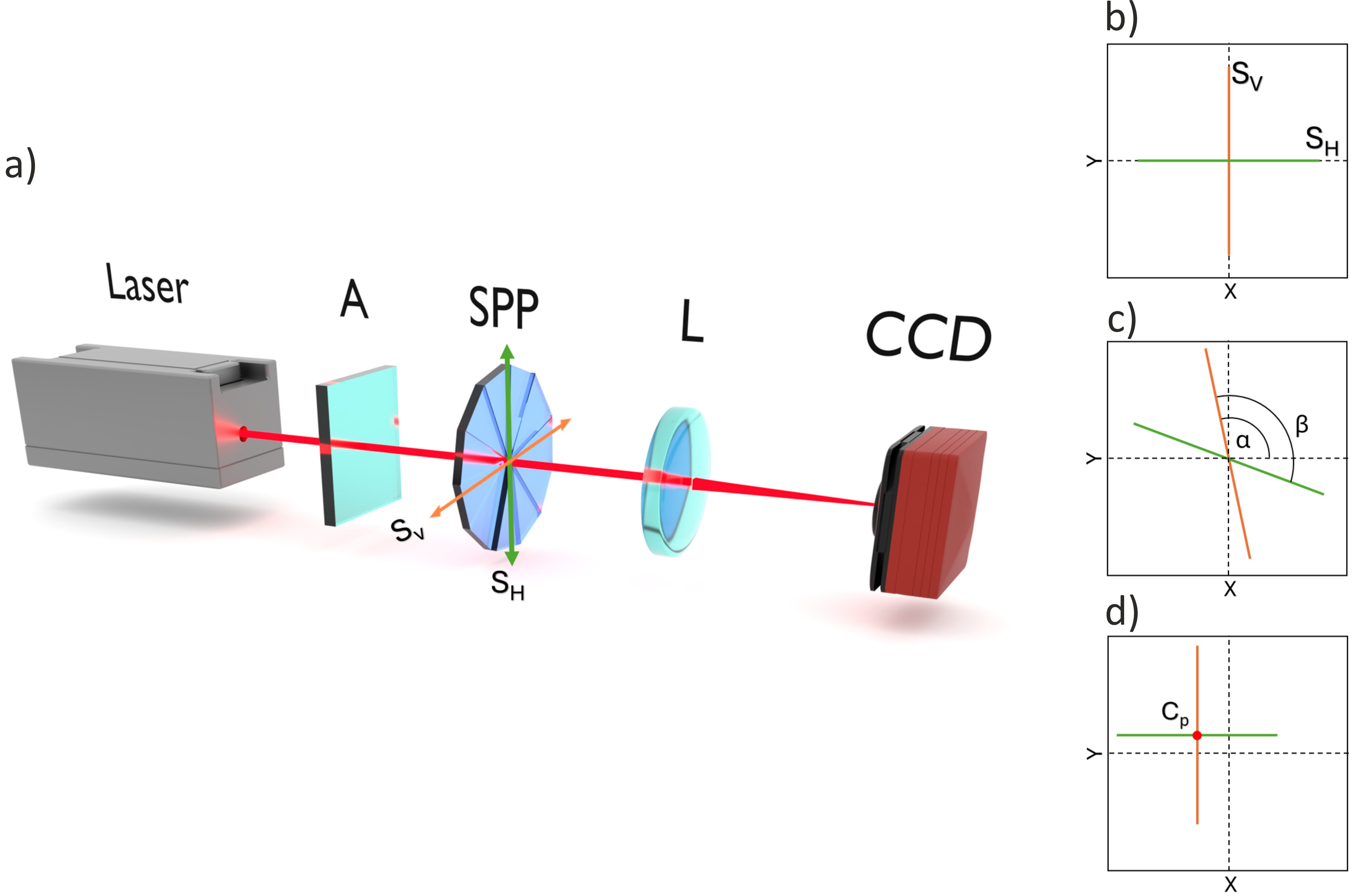}
\caption{(a) Numerical simulation model with SPP shift directions: green (\( S_H \), horizontal) and red (\( S_v \), vertical), both registered at the detector. 
(b) Example vortex trajectories showing perpendicularity between SPP shift and trajectory. 
(c) Inclined trajectories with angles \( \alpha \) (trajectory vs. x-axis) and \( \beta \) (between two trajectories). 
(d) Cross point \( C_p \) indicating trajectories intersection.}
\label{fig:setup_schematic}
\end{figure}

\begin{itemize}
    \item Laser source: A linearly polarized plane wave with a diameter of 250~\textmu m and a wavelength of 632.8~nm.
    \item Refractive element (A): Introduces the chosen aberration, with a diameter of 250~\textmu m.
    \item Spiral phase plate (SPP): Generates an optical vortex with a topological charge of \( m = 1 \), having the same diameter as element A.
    \item Focusing lens (L): Positioned at a distance of 100~\textmu m after the SPP.
    \item CCD detector: Placed at the back focal plane of L, 48.2~mm from the lens.
\end{itemize}

The only movable element in the model is the SPP, which is shifted in two perpendicular directions, vertical and horizontal, while maintaining a fixed z-distance along the optical axis. For each step of the SPP shift, the movement of the singular point is observed at the CCD detector. Then, the position of the singular point is detected and marked appropriately, which creates a single vortex trajectory after several shifts of the SPP. Trajectories at the detector plane are denoted by $S_v$ and $S_h$ labels, indicating the vortex trajectory. If not affected by any phase disturbance, $S_v$ and $S_h$ should create two perpendicular straight lines \cite{P-Masajada_2018, Plocinniczak_2016}, as schematically shown in Figure \ref{fig:setup_schematic}b. Moreover, the relation between the detected trajectory and the movement of the SPP is also perpendicular. Therefore, in this work, we refer to $S_v$ and $S_h$
as vortex trajectories, based on their orientation at the detector plane.

Both $S_v$ and $S_h$ are affected by any additional disturbances in the optical system. Figures \ref{fig:setup_schematic}c and \ref{fig:setup_schematic}d introduce additional parameters that characterize trajectory shape and correlate their deviations with specific aberrations.
The angle $\alpha$ represents the inclination between a vertical trajectory and the x-axis of the detector, while the angle $\beta$ denotes the relative angle between two trajectories. If both trajectories experience shifts, it can be quantified by the cross point $C_p$ defined as the intersection of the two trajectories.

\subsection{Detection of the singular point position}

The whole process of trajectory retrieval relies on the precise detection of the singular point. Although numerical simulations give access to the field's complex amplitude, which allows us to determine the singular point position directly  \cite{Gbur2018}. Intentionally, we proceeded with amplitude content only. This approach replicates an experimental scenario, allowing singularity tracking without the need for interference-based methods.
For that purpose, we implemented the Laguerre-Gaussian transform method \cite{Szatkowski2022}, which imposes the Laguerre-Gaussian filter \cite{Wang:06} on the registered intensity distribution, creating pseudo-complex amplitude and providing access to the singular point position. This approach is applied to both numerical and experimental data for consistency.

\subsection{Simulation parameters}

In the analysis, we considered primary monochromatic aberrations existing in regular optical systems, particularly astigmatism, defocus, and coma. Each aberration is represented by the Zernike polynomial coefficient $Z^l_n$ of the radial $l$ and azimuthal $n$ order, defined across a unitary aperture. In the presented numerical model, the system's aperture is determined by the aperture of (A) and set as 250 \textmu m. We retrieved both $S_v$ and $S_h$ trajectories for each aberration. These are the outcomes of two 60 \textmu m range scans per direction at the SPP plane. Each scan consists of 120 shifts with 0.5\textmu m step per scan. The process ends with the plot of two trajectories, respectively, detected at the focal plane, where the detector (CCD) is placed.

Further, we show how primary aberrations (defocus, astigmatism and coma) uniquely reflect optical trajectories and how to utilize this behavior in a simple auto-focusing algorithm.

\subsection{Simulation Results and Analysis}

In the first phase, the impact of single aberrations on vortex trajectories was examined. For that purpose, we introduced defocus $Z^0_2$, vertical astigmatism $Z^2_2$, horizontal $Z^1_3$ and vertical $Z^{-1}_3$ coma of different values, ranging from 0 to 0.1$\lambda$ with varying step. Two trajectories have been retrieved for each aberration value.

\begin{figure}[htbp]
\centering
\includegraphics[width=\textwidth]{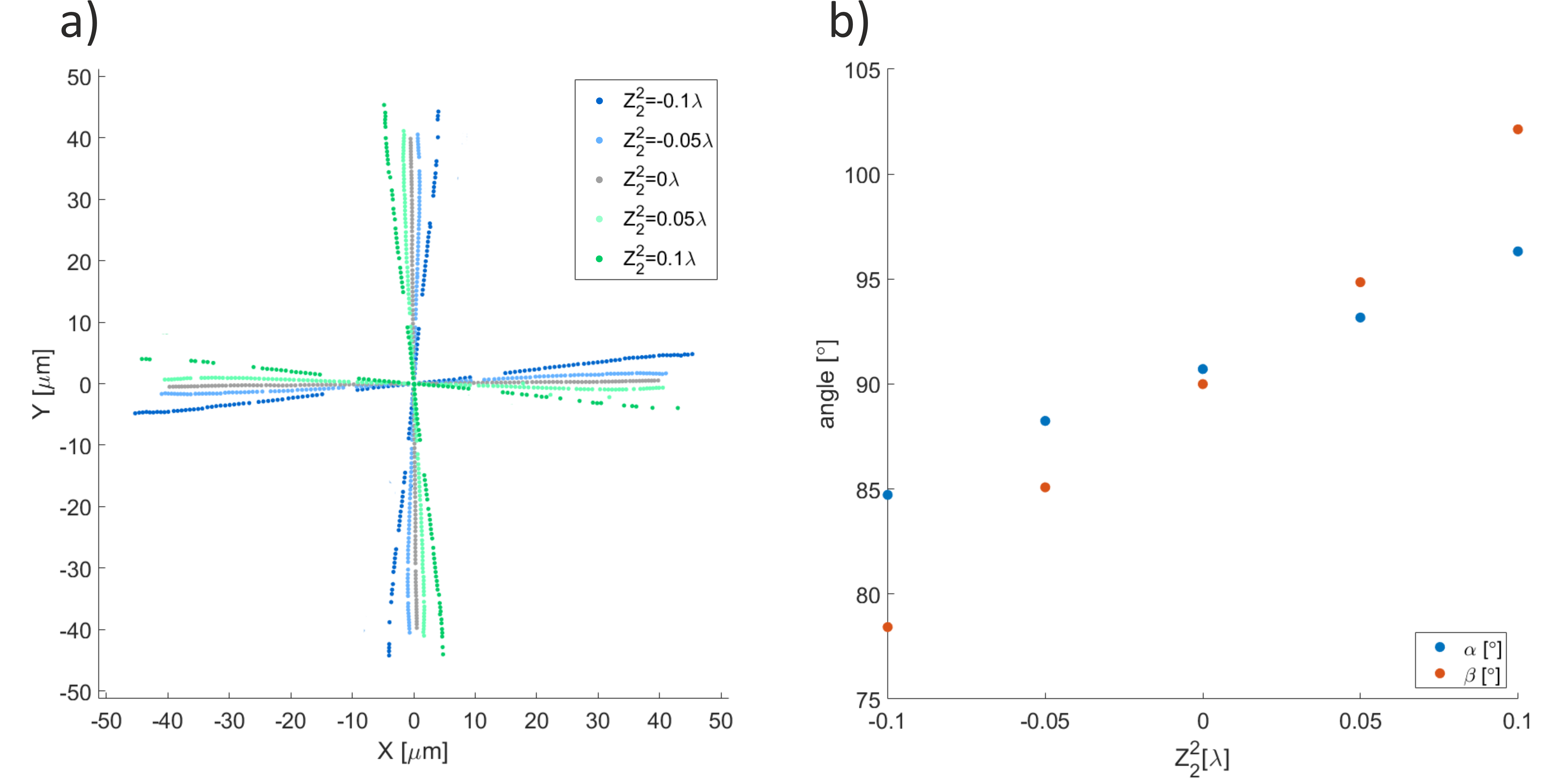}
\caption{a) Vortex trajectories retrieved for vertical astigmatism $Z^2_2$ b) Angles $\alpha$ (blue) and $\beta$ (red) in respect to $Z^2_2$ value.}
\label{fig:Astigmatism}
\end{figure}

As demonstrated experimentally in \cite{Szatkowski_2019} and confirmed by our simulations, astigmatism neither affects the cross-section point $C_p$ of two trajectories nor changes their linear shape (straight-line behavior is preserved) (Figure \ref{fig:Astigmatism}a). In the absence of astigmatism ($Z^2_2=0$), trajectories $S_v$ and $S_h$ remain almost perpendicular, with $\alpha=90.7^{\circ}$ and $\beta=90^{\circ}$. The slight $\alpha$ deviation from the ideal $90^{\circ}$ is caused by minor numerical artifacts in the trajectory data provided by the simulation software (missing data points). We found these deviations negligible as they do not affect the validity of the drawn conclusions.

When astigmatism is introduced, both angles increase for positive values of $Z^2_2$, reversing the trend for negative values of $Z^2_2$ (Figure \ref{fig:Astigmatism}b). The analysis of $\beta$ behavior shows that trajectories incline towards each other for the negative values of $Z^2_2$, while leading to larger $\beta$ angle values for positive $Z^2_2$. Although the analyzed range (-$0.1\lambda$, -$0.1\lambda$) of $Z^2_2$ value is limited, it already showed that the relation of aberration value and the change in both $\alpha$ and $\beta$ is not linear. The change of $\alpha$ is around $2.2^{\circ}$ and $\beta$ is equal $4.9^{\circ}$ for change the aberration value from $Z^2_2=0\lambda$ to $Z^2_2=0.05\lambda$. The increasing aberration value $Z^2_2$ from $0.05\lambda$ to $0.1\lambda$ causes change $\alpha$ of $4^{\circ}$ and $\beta$ of $7^{\circ}$.

Vortex trajectory behavior in this case is physically intuitive. Astigmatism creates two distinct focal planes: sagittal and tangential. We relate the focal plane to the perpendicularity of the detected vortex trajectory. When the observation plane remains fixed, astigmatism misaligns it with the focal planes. Consequently, the detected vortex trajectories are inclined rather than perpendicular.  

\begin{figure}[htbp]
\centering
\includegraphics[width=\textwidth]{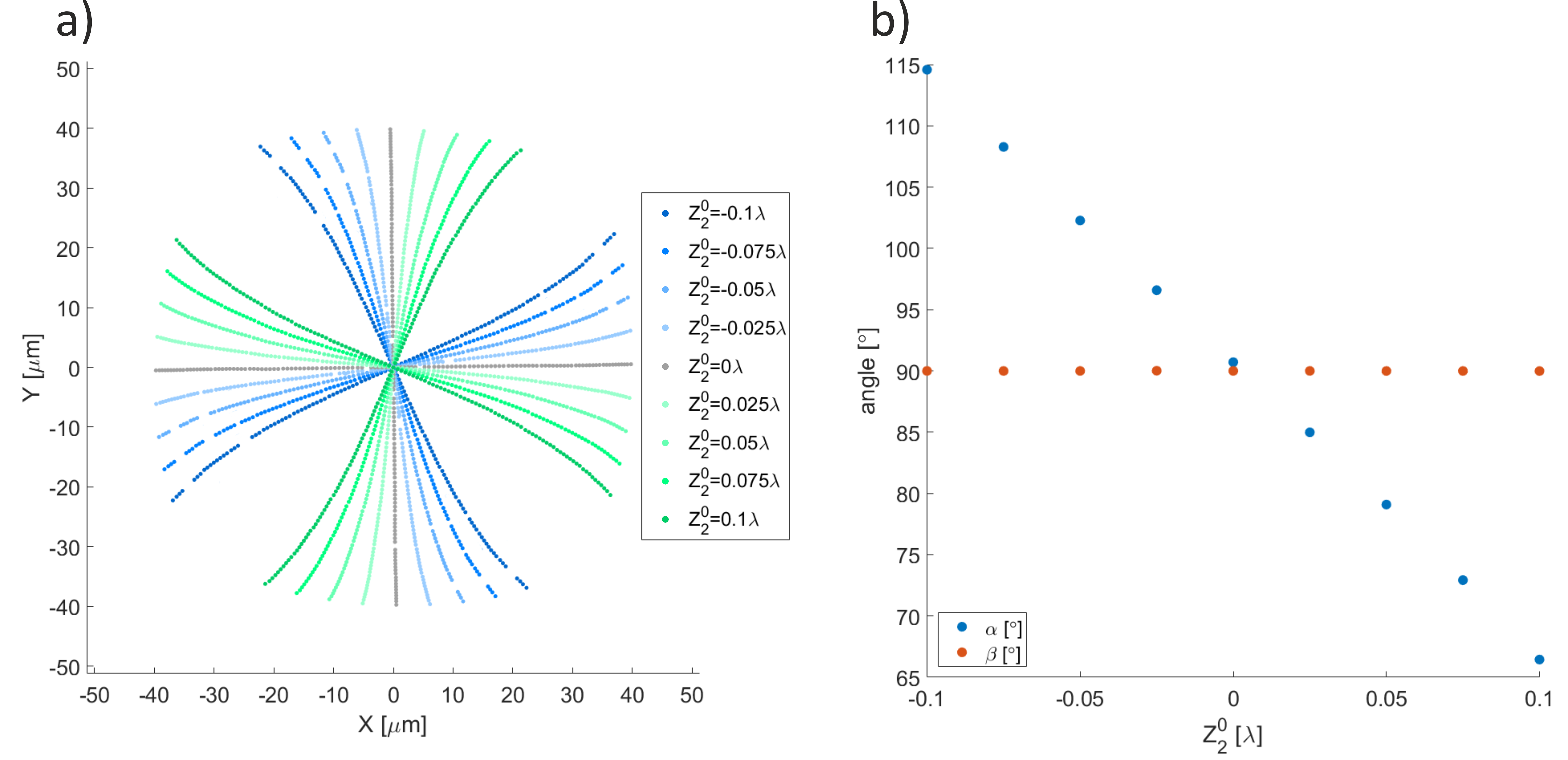}
\caption{a) Vortex trajectories retrieved for defocus $Z^0_2$ b) Angles $\alpha$ (blue) and $\beta$ (red) in respect to $Z^0_2$ value.}
\label{fig:Defocus}
\end{figure}

Following this behavior, one can shift the focal plane by introducing defocus. The focal plane is moved forward or backward along the optical axis, depending on the $Z^0_2$ sign. Since the position of the observation plane remains unchanged (initial focal plane of lens L), this will probe the vortex trajectory behavior in the system's focal plane ($ Z^0_2=0$) and its close vicinity ($Z^0_2\neq 0$) (Figure \ref{fig:Defocus}a). 

In the case of defocus, the angle between the two trajectories is preserved, and their perpendicularity is maintained. Both trajectories rotate in response to the introduced defocus and are sensitive to its sign (Figure~\ref{fig:Defocus}b). The angle $\beta$ oscillates around $90^{\circ}$ within the analyzed aberration range, a result of slight curvature in the trajectories when the off-axis vortex shift approaches the outer region of the beam. Despite these small fluctuations, a clear linear rotation behavior is observed in the angle $\alpha$, which decreases from $114^{\circ}$ to $66^{\circ}$ as the defocus coefficient $Z^0_2$ varies from $-0.1\lambda$ to $0.1\lambda$. In this range, the trajectories rotate by approximately $6^{\circ}$ for every $0.025\lambda$ change in $Z^0_2$, indicating high aberration sensitivity of the method—equivalent to a sensitivity of $\lambda/300$ per $1^{\circ}$ change.

\begin{figure}[htbp]
\centering
\includegraphics[width=\textwidth]{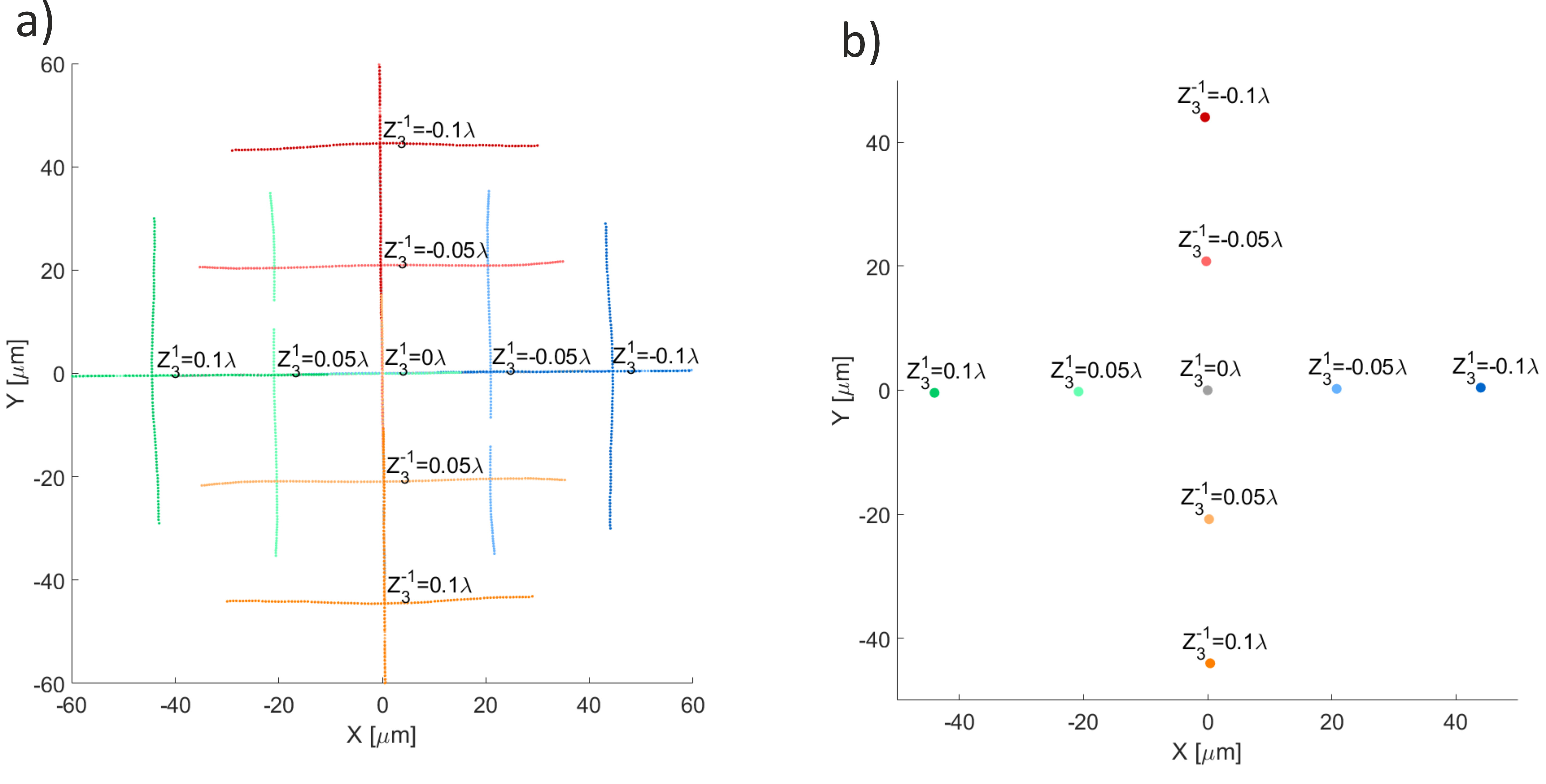}
\caption{a) Vortex trajectories retrieved for horizontal $Z^1_3$ and vertical $Z^{-1}_3$ denoting coma b) The aggregated positions of the cross-point $C_p$}
\label{fig:Coma}
\end{figure}

$Z^1_3$ and $Z^{-1}_3$, denoting horizontal and vertical comas, are the last aberration considered. The aggregated plot of retrieved trajectories for all analyzed cases is shown in Figure \ref{fig:Coma}a. The introduction of coma-type aberration shifts both trajectories without modifying angles $\alpha$ and $\beta$; trajectories remain unrotated and perpendicular, thus to analyze their impact, we highlighted the $C_p$ denoting the cross-point of trajectories (Figure \ref{fig:Coma}b). Its position is determined by the type of coma, value of the aberration and its sign. We observed the vertical movement for $Z^{-1}_3$, and, analogically, a horizontal one for $Z^1_3$. Each $C_p$ changed it's position by 14.6 $\mu m$ per $Z^l_2=0.025\lambda$ with an accuracy of less than 0.4 $\mu m$.

The vortex trajectory response is unique for each of these aberrations. Astigmatism inclines two trajectories, defocus rotates them, and coma moves the cross-point without affecting the shape of trajectories. These results can serve as a good and objective indicator of the aberration existing in the optical setup and can be efficiently applied in basic setup alignment. 

The following paragraph shows how these results could be applied as an autofocusing tool that does not rely on image analysis but uses objective parameters, such as optical vortex trajectory. 

\subsection{Off-axis dark ray}

To utilize vortex trajectory as an autofocusing tool, we leverage the effect of defocus on the position of the singular point. Figure \ref{fig:Defocus}a showed the result of trajectories observed at the fixed observation plane, under manually introduced defocus. Here, we present an alternative concept that examines dark ray evolution - the trace of the singular point along the optical axis. 

\begin{figure} [htbp]
    \centering
    \includegraphics[width=\linewidth]{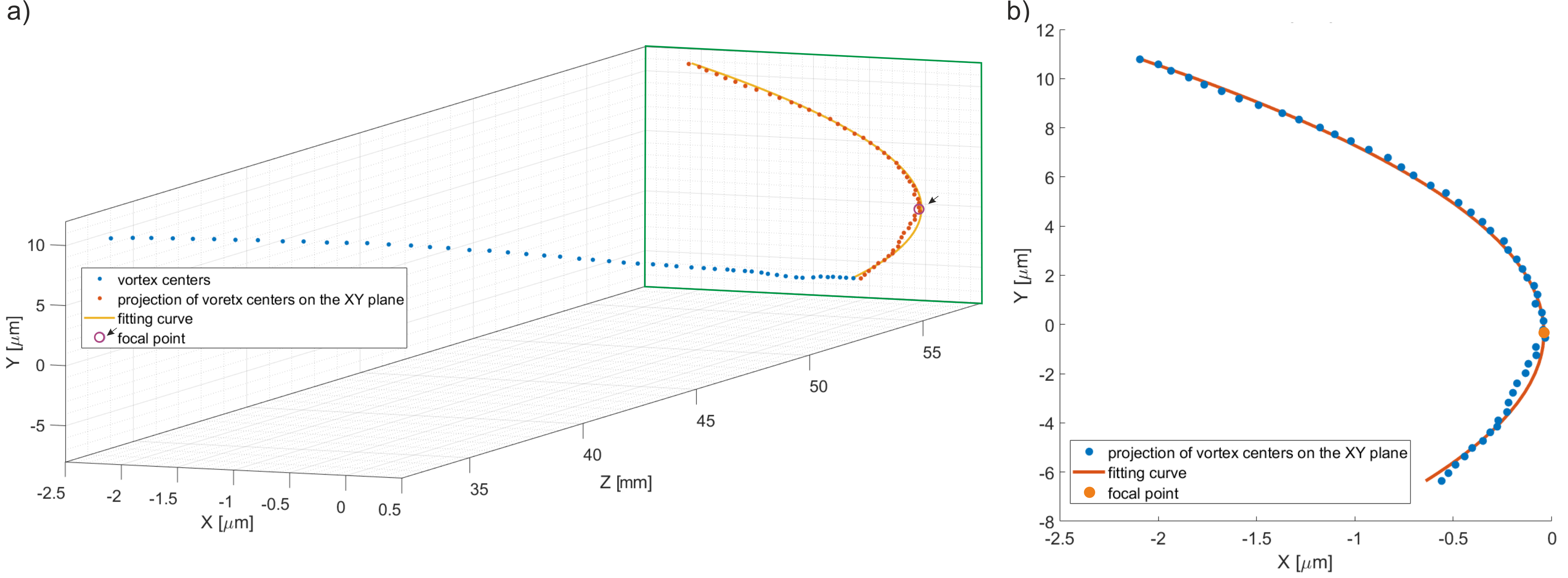}
    \caption{a) Evolution of the off-axis ($\Delta S_H=40 \mu m$) singular point along the optical axis Z. The green area represents the projection of all singularity positions onto the XY plane, further magnified in b). The singularity positions form a parabolic trajectory, where the local extremum corresponds to the focal plane of lens L, denoted as the focal point on the fitted curve (orange).}
    \label{fig:Dark_ray}
\end{figure}

Out of the positions shown in Figure \ref{fig:Defocus}a, we select a single off-axis vortex position, $\Delta S_H=40 \mu m$, and analyze its evolution by shifting the detector along the optical axis over the range $z = (15 \text{ mm}:0.5 \text{ mm}:70 \text{ mm})$, measured from the focusing lens L. The singular point is tracked by recording its position at each step of $z$. An example of the resulting dark ray is shown in Figure \ref{fig:Dark_ray}a.

As shown in Figure \ref{fig:Dark_ray}b, the projection of the dark ray onto the XY plane forms a parabolic shape $f(z)$, representing the trace of the singular point as a function of finite $N$ detector positions $z$. The extremum of this parabola corresponds to the plane $z_{\text{ext}}$ where the vortex trajectory is perpendicular to the shift of the vortex-generating element, thus, we look for the condition, where:

\[
z_{\text{ext}} = \{z_i : f'(z_i) \cdot f'(z_{i+1}) < 0\} \quad 2 \leq i \leq N-1
\]

Perceeded by calculating first derivative of $f'(z_i)$:

\[
f'(z_i) \approx \frac{f(z_{i+1}) - f(z_{i-1})}{z_{i+1} - z_{i-1}}
\]

Alternatively, one can define the error margin $\epsilon$ and determine $ z_{\text{ext}}$ as:

\[
f'(z_{\text{ext}}) \leq \epsilon
\]

Or interpolate discrete $f(z)$ and determine its extremum analytically. 

Following the analytical study in \cite{Augustyniak_2012, Masajada2013}, this perpendicularity condition is satisfied at the plane where the wavefront curvature radius approaches infinity, which typically coincides with the beam waist in an ideal optical system. In the numerical example (Figure \ref{fig:Dark_ray}), the calculated $ z_{\text{ext}}=48.2 mm$, which matches the back focal plane of the lens L. As stated in \cite{Augustyniak_2012, Masajada2013}, the off-axis shift $\Delta S_n$ should not exceed $20\%$ of the beam diameter, where $n$ denotes either horizontal ($H$) or vertical ($V$) direction. In our model, the chosen $\Delta S_H=40 \mu m$ falls below this limit (beam diameter $=250 \mu m$).

In S1, we elaborate on the analogies between Figure \ref{fig:Defocus} and Figure \ref{fig:Dark_ray}, as they complement one another. In the following section, we experimentally employ this approach to identify the system's focal plane.

\section{Experiment}

\begin{figure}[htbp]
\centering
\includegraphics[width=350px]{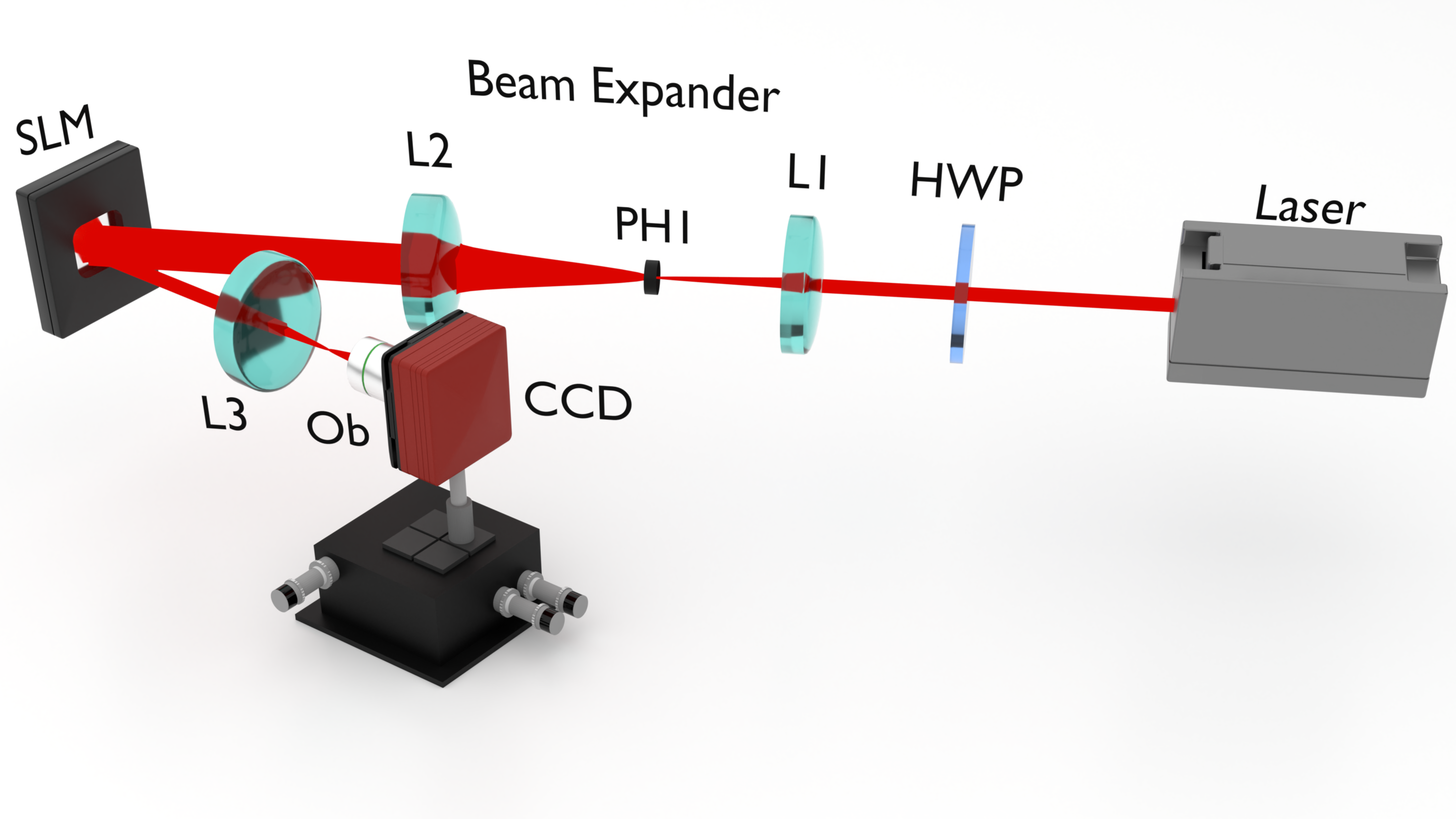}
\caption{Scheme of the experimental setup. A Spatial Light Modulator works as a vortex-generating element, displaying both on-axis and off-axis vortices. The CCD camera, together with imaging objective Ob are mounted on the motorized plane that controls their position on the optical axis.}
\label{fig:setup_experimental}
\end{figure}

To demonstrate the performance of vortex trajectories for determining the system's focal plane, the experimental setup presented in Figure \ref{fig:setup_experimental} has been proposed. The expanded laser beam illuminates the Holoeye LC-R2500 SLM, which operates in phase-only mode due to the proper orientation of a half-wave plate that rotates the linear polarization. The SLM generates a Laguerre-Gaussian mode, which is then focused by lens L3 and imaged onto the CCD camera through objective Ob. The theoretical back focal plane of L3 is $f_{L3}= 100 mm$ for a plane wave illumination, which has been achieved by the phase correction of the SLM. The objective and the camera are mounted on a motorized stage that shifts them along the optical axis. The entire range of the observation plane shift (objective and CCD camera) along the optical axis was set to $z = (99 mm : 0.021 mm : 101 mm)$ with a total number of 100 steps.

Here, two positions of the Laguerre-Gaussian beam are considered. On-axis (Laguerre-Gaussian generating hologram is centered with the illuminating beam), and off-axis, where the hologram is mathematically translated so that $\Delta S_n \neq 0$. The on-axis position serves as a reference for positioning the beam, in case it shifts due to a tilt of any of the optical elements. Therefore, the measurement protocol can be reduced to the following steps:

\begin{enumerate}
    \item Set the initial detector position at \( z_i \).
    \item Capture an image of the on-axis Laguerre-Gaussian beam and determine \( S_n^{\text{on}}(z_i) \).
    \item Capture an image of the off-axis Laguerre-Gaussian beam and determine \( S_n^{\text{off}}(z_i) \).
    \item Move the detector to the next position \( z_{i+1} \).
    \item Repeat steps 2–5 until the final detector position is reached.
\end{enumerate}
 
Using the Laguerre-Gaussian transform localization algorithm \cite{Szatkowski2022}, we end up with on-axis \( S_n^{\text{on}} \) and off-axis \( S_n^{\text{off}} \) positions of the singular point for a single $z_{i}$. Then, the dark ray $f(z_i)$ is formed by:
\[
f(z_i) = S_n^{\text{off}}(z_i) - S_n^{\text{on}}(z_i)
\]

\begin{figure} [htbp]
    \centering
    \includegraphics[width=\linewidth]{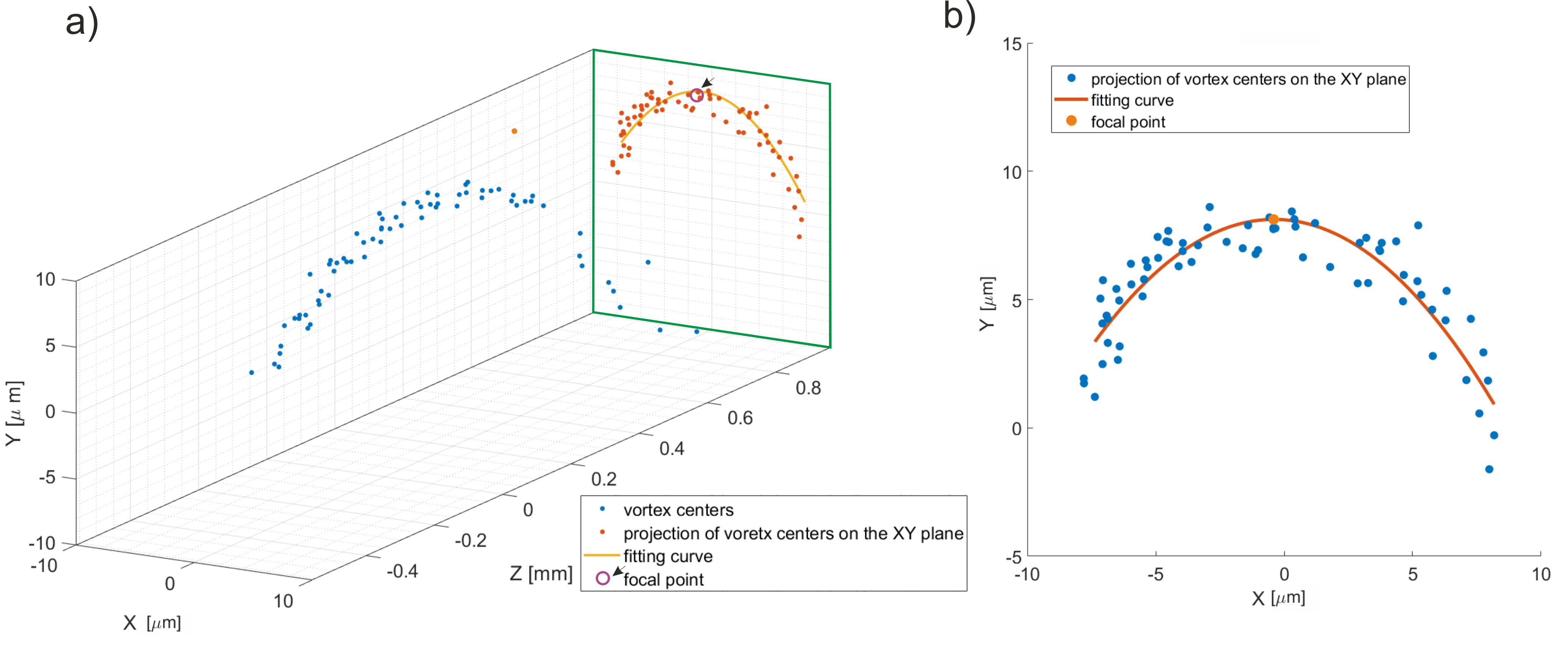}
    \caption{a) Experimental evolution of the off-axis ($\Delta S_v=-50 \mu m$) singular point along the optical axis Z. The green area represents the projection of all singularity positions onto the XY plane, further magnified in b) The local extremum corresponds to the focal plane of lens L3, denoted as the focal point on the fitted curve (orange).}
    \label{dark_ray_experimental}
\end{figure}

This ensures that the algorithm is not affected by optical misalignment, which could introduce additional beam tilt. Figure \ref{dark_ray_experimental}a shows the dark ray, while Figure \ref{dark_ray_experimental}b shows its projection on the XY detector plane. The off-axis shift was set to $\Delta S_V=-50 \mu m$, and for the chosen off-axis shift, the detected extrema points out to the focal plane position equal to $100.098 mm$, which matches the theoretical $f_{L3}=100 mm$. 

The precision of the algorithm depends on the value of the $\Delta S_n$, which directly impacts the curviness of the dark ray projection on the XY plane. In the experimental verification, the beam diameter at the SLM was equal to $10.45 mm$ and for the entire range $\Delta S_V=[-30 \mu m, -70 \mu m] $, the detected back focal plane position did not exceed $100.180 mm$ (See S2). Thus, we estimate the precision of this method to be $< 0.2 mm$.

\section{Conclusion}

The main goal of this work is to examine how primary optical aberrations affect singular point evolution. In the proposed numerical model, we analyzed the effects of defocus, coma, and astigmatism, establishing a direct link between specific optical vortex trajectory features and aberration type and value. These results show that each aberration can be linked with a unique trajectory behavior.

The proposed methodology to evaluate optical aberrations does not involve interferometry and relies only on the behavior of the singular point position, which can be precisely calculated. The localization algorithm uses the a priori information about the vortex existence and works even in the low light regime \cite{Szatkowski2022}. 

Although complete wavefront retrieval requires further analytical study, the potential benefits include high sensitivity without additional wavefront sensing devices, as long as the vortex-generating element is already part of the setup, building on the popularity of Spatial Light Modulators or Digital Micromirror Devices.

We demonstrated the immediate application of the proposed methodology as an auto-focusing algorithm that identifies the focal plane by evaluating the position of the off-axis singular point due to the shift in the observation plane. This approach is easily automated and is based on the objective metric: the perpendicular relationship between the off-axis singular point position and the vortex-generating element shift.

Lastly, this work showed how to utilize the dark ray to measure optical systems' aberration. Future studies should explore the relationship between trajectory parameters and the superposition of multiple aberrations. This could further advance the wavefront sensing with phase singularities.

\section*{Acknowledgments}
This research was funded in whole by National Centre for Research and Development, 0230/L-13/2022.

\section*{Author Contributions}
\textbf{A.K. Korzeniewska}: Conceptualization (supporting); Methodology (equal); Investigation (equal); Validation (lead); Writing – review \& editing (supporting).
\textbf{M. \L{}ukowicz}: Investigation (equal); Validation (supporting).
\textbf{K. Kalinowski}: Methodology (supporting); Software (lead).
\textbf{K. Gemza}: Methodology (equal); Software (supporting); Validation (supporting).
\textbf{M. Szatkowski}: Conceptualization (lead); Funding acquisition (lead); Supervision (lead); Validation (supporting); Writing – original draft (lead).

\section*{Disclosures}
The authors declare no conflicts of interest.

\section*{Data Availability}
The data supporting this study's findings are available from the corresponding author upon reasonable request.

\bibliographystyle{unsrt}
\bibliography{references}

\begin{thebibliography}{10}

\bibitem{Nye1974}
John~Frederick Nye and Michael~V. Berry.
\newblock Dislocations in wave trains.
\newblock {\em Proceedings of the Royal Society of London. A. Mathematical and Physical Sciences}, 336:165 -- 190, 1974.

\bibitem{Basistiy1995}
I.~V. Basistiy, M.~S. Soskin, and M.~V. Vasnetsov.
\newblock Optical wavefront dislocations and their properties.
\newblock {\em Optics Communications}, 119:604--612, 1995.

\bibitem{Bazhenov1992}
V.Yu. Bazhenov, M.S. Soskin, and M.V. Vasnetsov.
\newblock Screw dislocations in light wavefronts.
\newblock {\em Journal of Modern Optics}, 39:985--990, 5 1992.

\bibitem{Freund1999}
Isaac Freund.
\newblock Critical point explosions in two-dimensional wave fields, 1999.

\bibitem{Dennis2012}
Mark~R. Dennis and Jörg~B. Götte.
\newblock Topological aberration of optical vortex beams: Determining dielectric interfaces by optical singularity shifts.
\newblock {\em Physical Review Letters}, 109, 10 2012.

\bibitem{Roux2003}
Filippus~S. Roux.
\newblock {Paraxial modal analysis technique for optical vortex trajectories}.
\newblock {\em Journal of the Optical Society of America B}, 20(7):1575, 2003.

\bibitem{Roux2004}
Filippus~S. Roux.
\newblock Spatial evolution of the morphology of an optical vortex dipole.
\newblock {\em Optics Communications}, 236:433--440, 6 2004.

\bibitem{Bekshaev2017}
Aleksandr Bekshaev, Aleksey Chernykh, Anna Khoroshun, and Lidiya Mikhaylovskaya.
\newblock {Singular skeleton evolution and topological reactions in edge-diffracted circular optical-vortex beams}.
\newblock {\em Optics Communications}, 397:72--83, aug 2017.

\bibitem{Khoroshun2017}
Anna Khoroshun, Aleksey Chernykh, Julia Kirichenko, Oleksandr Ryazantsev, and Aleksandr Bekshaev.
\newblock Singular skeleton of a laguerre–gaussian beam transformed by the double-phase-ramp converter.
\newblock {\em Applied Optics}, 56:3428, 2017.

\bibitem{Khoroshun2020}
A.~Khoroshun, A.~Ryazantsev, O.~Ryazantsev, S.~Sato, Y.~Kozawa, J.~Masajada, A.~Popiołek-Masajada, M.~Szatkowski, A.~Chernykh, and A.~Bekshaev.
\newblock Formation of an optical field with regular singular-skeleton structure by the double-phase-ramp converter.
\newblock {\em Journal of Optics (United Kingdom)}, 22:9, 2020.

\bibitem{Ferrando2013}
A.~Ferrando and M.~A. García-March.
\newblock Theory for the control of dark rays by means of discrete symmetry diffractive elements.
\newblock {\em Journal of Optics (United Kingdom)}, 15, 2013.

\bibitem{Leach2005}
J.~Leach, M.~R. Dennis, J.~Courtial, and M.~J. Padgett.
\newblock {Vortex knots in light}.
\newblock {\em New Journal of Physics}, 7, feb 2005.

\bibitem{Dennis2010}
Mark~R. Dennis, Robert~P. King, Barry Jack, Kevin Oholleran, and Miles~J. Padgett.
\newblock Isolated optical vortex knots.
\newblock {\em Nature Physics}, 6:118--121, 2010.

\bibitem{Dehghan2023}
Nazanin Dehghan, Alessio D'Errico, Tareq Jaouni, and Ebrahim Karimi.
\newblock {Effects of aberrations on 3D optical topologies}.
\newblock {\em Communications Physics}, 6(1), dec 2023.

\bibitem{Larocque2020}
Hugo Larocque, Alessio D'Errico, Manuel~F. Ferrer-Garcia, Avishy Carmi, Eliahu Cohen, and Ebrahim Karimi.
\newblock {Optical framed knots as information carriers}.
\newblock {\em Nature Communications}, 11(1), dec 2020.

\bibitem{Kong2022}
Ling~Jun Kong, Weixuan Zhang, Peng Li, Xuyue Guo, Jingfeng Zhang, Furong Zhang, Jianlin Zhao, and Xiangdong Zhang.
\newblock High capacity topological coding based on nested vortex knots and links.
\newblock {\em Nature Communications}, 13, 12 2022.

\bibitem{Ortega2019}
A.~Balbuena Ortega, S.~Bucio-Pacheco, S.~Lopez-Huidobro, L.~Perez-Garcia, F.~J. Poveda-Cuevas, J.~A. Seman, A.~V. Arzola, and K.~Volke-Sepúlveda.
\newblock Creation of optical speckle by randomizing a vortex-lattice.
\newblock {\em Optics Express}, 27:4105, 2019.

\bibitem{Karen_Volke_2022}
Argelia~Balbuena Ortega, Esteban Vélez-Juárez, and Karen Volke-Sepúlveda.
\newblock Structure transitions in arrays of point-vortices upon free space propagation.
\newblock {\em Journal of Optics (United Kingdom)}, 24, 12 2022.

\bibitem{Ferrando2023}
Albert Ferrando, Agnieszka Popiołek-Masajada, Jan Masajada, Raman Markevich, and Anna Khoroshun.
\newblock Vortex-antivortex pair control in quadrupole gaussian beams.
\newblock {\em Optics Express}, 31:23444, 7 2023.

\bibitem{Augustyniak_2012}
Ireneusz Augustyniak, Agnieszka Popio{\l}ek-Masajada, Jan Masajada, and S{\l}awomir Drobczy\'{n}ski.
\newblock New scanning technique for the optical vortex microscope.
\newblock {\em Appl. Opt.}, 51(10):C117--C124, Apr 2012.

\bibitem{P-Masajada_2018}
Agnieszka Popiolek-Masajada, Jan Masajada, and Mateusz Szatkowski.
\newblock Internal scanning method as unique imaging method of optical vortex scanning microscope.
\newblock {\em Optics and Lasers in Engineering}, 105:201--208, 02 2018.

\bibitem{Plocinniczak_2016}
{\L}ukasz P{\l}ocinniczak, Agnieszka Popio{\l}ek-Masajada, Jan Masajada, and Mateusz Szatkowski.
\newblock Analytical model of the optical vortex microscope.
\newblock {\em Appl. Opt.}, 55(12):B20--B27, Apr 2016.

\bibitem{He2022}
Chao He, Yijie Shen, and Andrew Forbes.
\newblock {Towards higher-dimensional structured light}.
\newblock {\em Light: Science and Applications}, 11(1), dec 2022.

\bibitem{Berry_2019}
Michael Berry, Nikolay Zheludev, Yakir Aharonov, Fabrizio Colombo, Irene Sabadini, Daniele~C Struppa, Jeff Tollaksen, Edward T~F Rogers, Fei Qin, Minghui Hong, Xiangang Luo, Roei Remez, Ady Arie, Jörg~B Götte, Mark~R Dennis, Alex M~H Wong, George~V Eleftheriades, Yaniv Eliezer, Alon Bahabad, Gang Chen, Zhongquan Wen, Gaofeng Liang, Chenglong Hao, C-W Qiu, Achim Kempf, Eytan Katzav, and Moshe Schwartz.
\newblock Roadmap on superoscillations.
\newblock {\em Journal of Optics}, 21(5):053002, apr 2019.

\bibitem{Gbur2018}
Greg Gbur.
\newblock Using superoscillations for superresolved imaging and subwavelength focusing.
\newblock {\em Nanophotonics}, 8:205--225, 2018.

\bibitem{Chen2019}
Gang Chen, Zhong~Quan Wen, and Cheng~Wei Qiu.
\newblock Superoscillation: from physics to optical applications.
\newblock {\em Light: Science and Applications}, 8, 2019.

\bibitem{Szatkowski_2019}
Mateusz Szatkowski, Agnieszka {Popiołek Masajada}, and Jan Masajada.
\newblock Optical vortex trajectory as a merit function for spatial light modulator correction.
\newblock {\em Optics and Lasers in Engineering}, 118:1--6, 2019.

\bibitem{Szatkowski2022}
Mateusz Szatkowski, Emilia Burnecka, Hanna Dy{\l}a, and Jan Masajada.
\newblock Optical vortex tracking algorithm based on the laguerre-gaussian transform.
\newblock {\em Opt. Express}, 30(10):17451--17464, May 2022.

\bibitem{Wang:06}
Wei Wang, Tomoaki Yokozeki, Reika Ishijima, Mitsuo Takeda, and Steen~G. Hanson.
\newblock Optical vortex metrology based on the core structures of phase singularities in laguerre-gauss transform of a speckle pattern.
\newblock {\em Opt. Express}, 14(22):10195--10206, Oct 2006.

\bibitem{Masajada2013}
J.~Masajada, I.~Augustyniak, and A.~Popiołek-Masajada.
\newblock Optical vortex dynamics induced by vortex lens shift-optical system error analysis.
\newblock {\em Journal of Optics (United Kingdom)}, 15:044031, 2013.

\end{thebibliography}

\section*{Supplemental document}
\label{Supplemental_document}

\noindent\textbf{S1 Complementary measurement paths for vortex trajectory and dark ray analysis}
\label{sec:supplement_S1}

To provide broader context, Figure~\ref{Supplement_1} illustrates the two alternative approaches presented in this work. Path 1 assumes that the CCD position is fixed along the optical axis \( Z \) and aligned with the back focal plane of the focusing lens \( L \). In part 1.1, the vortex position is calculated for each step of the SPP shift (here, \( S_V \)) and then plotted (part 1.2), forming a single trajectory corresponding to a specific defocus value set by the aberration-generating element \( A \). By modifying the strength of the aberration and repeating the SPP scan, we obtain the final plot (part 1.3), which directly leads to Figure~\ref{fig:Defocus}.

Analogously, we can reverse this concept in Path 2, where the SPP is shifted off-axis to a fixed position \( \Delta S_H \) to generate an off-axis optical vortex, while the CCD is translated along the optical axis \( Z \). The resulting sequence of recorded images is shown in part 2.1, with the corresponding dark ray, gathering all singular point positions, depicted in part 2.2, along with its projection onto the \( XY \) plane—thus forming Figure~\ref{fig:Dark_ray}. This provides a direct analogy to Figure~\ref{fig:Defocus}, as translating the CCD is equivalent to modifying the defocus value. Exemplary positions along this trajectory are indicated by red dots.

\renewcommand{\thefigure}{S\arabic{figure}}
\setcounter{figure}{0}
\begin{figure}
    \centering
    \includegraphics[width=\linewidth]{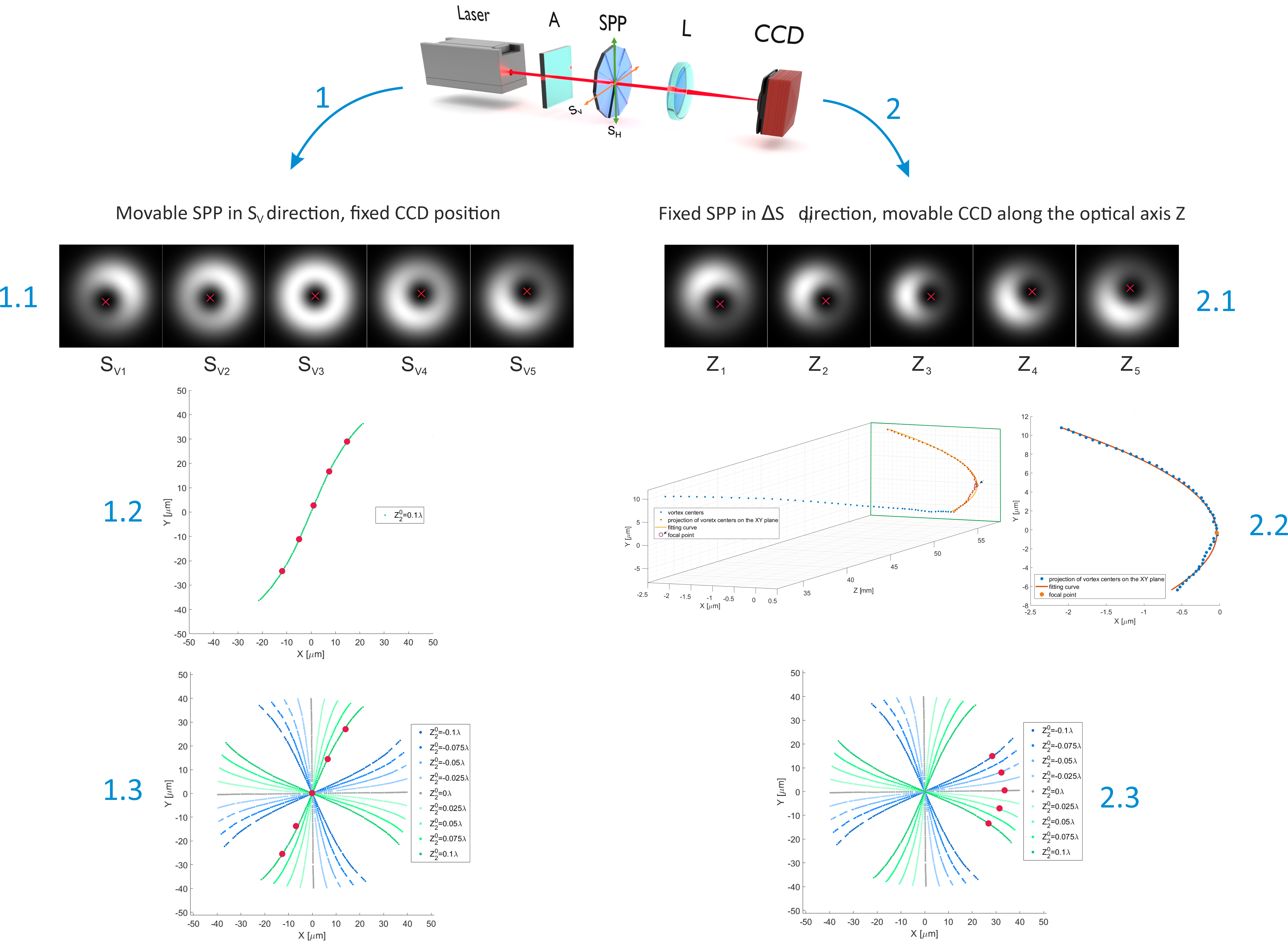}
    \caption{Schematic representation of two alternative measurement paths used for vortex trajectory and dark ray analysis}
    \label{Supplement_1}
\end{figure}

\noindent\textbf{S2 Dependence of the vortex generating element shift and accuracy of the algorithm}
\label{Section_suplement_2}

\begin{figure}[H]
    \centering
    \includegraphics[width=\linewidth]{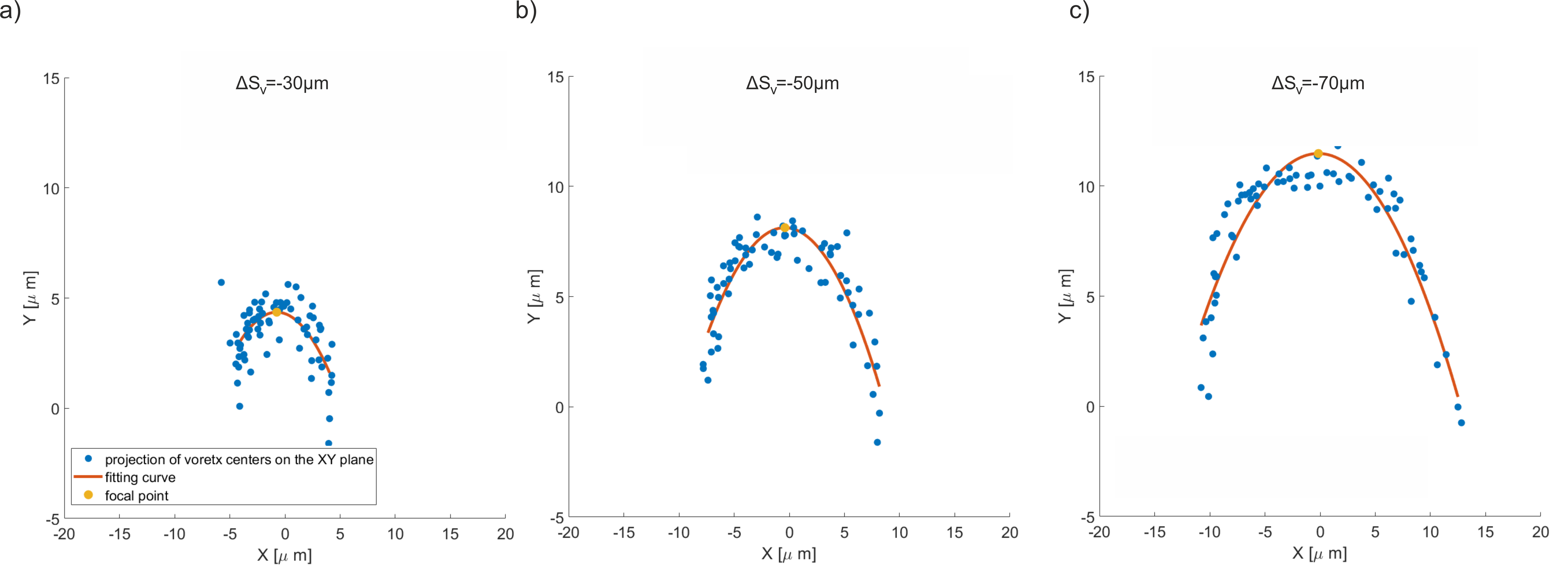}
    \caption{Projection of all singularity positions
onto the XY plane for different vertical shifts of the vortex generating element, where a) $\Delta S_v=-30 \mu m$, b) $\Delta S_v=-50 \mu m$ (same data as shown in Figure\ref{dark_ray_experimental}b), c) $\Delta S_v=-70 \mu m$
}
    \label{Supplement_2}
\end{figure}

Following the analytical analysis of vortex trajectories~\cite{Augustyniak_2012, Masajada2013}, the off-axis shift (\(\Delta S_n\)) should not exceed \(20\%\) of the beam diameter, to avoid annihilation of the singular point by the outer part of the beam. The choice of the \(\Delta S_n\) value is crucial and impacts the accuracy of the focal plane determination. In Figure~\ref{Supplement_2}, we present additional experimental results for different values of \(\Delta S_v\) together with calculated extrema, which indicate the back focal plane position. Calculated back focal plane values are further shown in Table~\ref{tab:shift_comparison}. The theoretical back focal plane of the system is set to 100~mm, which shows good agreement with experimental results. 

While the shift of the vortex-generating element ($\Delta S_v$) does influence the calculated focal plane position, its effect remains minor as long as the shift stays within the analytically recommended range. The beam diameter at the SLM plane was equal to $10.45 mm$, and the chosen $Delta S_v$, despite being less than $1\%$ of the beam size, led to correct back focal plane predictions. These results confirm the robustness of the method for accurate back focal plane localization.

\renewcommand{\thetable}{S\arabic{table}}
\setcounter{table}{0}  
\begin{table}[htbp]
\centering
\begin{tabular}{|l|l|}
\hline
$\boldsymbol{\Delta S_V} \boldsymbol{[\mu m]}$ & \textbf{Calculated back focal plane [$mm$]} \\
\hline
-30 & 100.173 \\
\hline
-50 & 100.098 \\
\hline
-70 & 100.043 \\
\hline
\end{tabular}
\caption{Experimentally determined back focal plane for different values of $\Delta S_V$}
\label{tab:shift_comparison}
\end{table}

\end{document}